\documentclass[superscriptaddress,aps,prl,twocolumn]{revtex4}

\def\beq{\begin{equation}}
\def\eeq{\end{equation}}
\def\bea{\begin{eqnarray}}
\def\eea{\end{eqnarray}}

\usepackage{graphicx}

\begin{document}

\title{Ground State Properties of Fermi Gases in the Strongly Interacting Regime}
\stepcounter{mpfootnote}
\author{S. Y. Chang}
\author{V. R. Pandharipande}
\address { Department of Physics,  
 University of Illinois at Urbana-Champaign,
       1110 W. Green St., Urbana, IL 61801, U.S.A.}

\date{\today}
\begin{abstract}
The ground state energies and pairing gaps in dilute superfluid Fermi gases have now been
calculated with the quantum Monte Carlo method without detailed knowledge of their wave functions.
However, such knowledge is essential to predict other properties of these gases such as density
matrices and pair distribution functions. We present a new and simple method to optimize the wave
functions of quantum fluids using Green's function Monte Carlo method.  It is used to calculate
 the pair distribution functions and potential energies of Fermi gases over the entire regime from 
atomic Bardeen-Cooper-Schrieffer superfluid to molecular Bose-Einstein condensation, spanned as 
the interaction strength is varied.  PACS: 03.75.Ss, 21.65.+f, 02.70.Ss
\end{abstract}
\maketitle

Recent progress in experimental \cite{demarco1999, ohara2002, roberts2001,regal2003,
regal2004, bartenstein2004} and theoretical methods \cite{randeria95, carlson2003,
chang2004, astrakharchik2004} has generated great interest in the properties of dilute Fermi superfluid gases.
Such gases are also of interest in studies of astrophysical objects such as neutron stars \cite{pethick95}
and in nuclear physics \cite{dean2003}.

The Hamiltonian of these gases has the standard form
\beq
{\cal H}=-\frac{\hbar^2}{2m} \sum_i \nabla^2_i + \sum_{i<j} v(r_{ij})~.
\eeq
The range of the interatomic potential $v(r_{ij})$ is much smaller than the interparticle spacing in 
the gas, and only the s-wave scattering length $a$ of the interaction is relevent.  Weak 
attractive interactions have a small negative $a$ which increases in magnitude as the 
interaction gets stronger.  The $a \rightarrow -\infty$ as we approach the bound molecular state. 
On further increase of the interaction strength, $a$ goes discontinuosly to $+\infty$ and then
smoothly to 0 as the molecule gets more tightly bound. 

 Usually, the dimensionless quantity $1/ak_F$ is used to characterize the gas. 
When the interaction is weak and attractive, $1/ak_F \rightarrow - \infty$, 
and we have a BCS superfluid gas with gap
$\Delta \sim e^{\pi/(2ak_F)}$ (BCS regime).  It has
$1/ak_F << 0$, $\Delta <<$ the energy per particle $E_0/N$ which is positive and
less than the Fermi gas energy $E_{FG} = \frac{3}{5} \frac{\hbar^2 k_F^2}{2m}$.
When the interaction is strong, $1/ak_F >> 0 $, we have tightly bound molecules with energy $E_{mol}$, and
$E_0/N \approx  E_{mol}/2 \approx -\Delta$. The Bose molecules are condensed in the zero
momentum state (BEC regime). In the intermediate regime ($-0.5 \lesssim 1/ak_F \lesssim 0.5$) we seem to have
a smooth transition or crossover from BCS superfluid to BEC. 

The problem of calculating the ground state energies and pairing gaps of superfluid 
gases has been solved with the fixed node Green's
function Monte Carlo (FN-GFMC) method  \cite{carlson2003, chang2004,astrakharchik2004}.
To begin with, we review this method and the problem it faces for computing observables 
other than energies.  

In FN-GFMC a trial wave function $\Psi_T({\bf R})$ is evolved in imaginary time $\tau$ 
with the fixed node constraint \cite{anderson75}
\beq
\Psi(\tau , {\bf R})  =  \left[e^{-\tau ({\cal{H}} - E_T)}\right]_{FN} ~ \Psi_T({\bf R})~.
\label{t_evol}
\eeq
We use $ {\bf R} = {\bf r}_1,{\bf r}_2,\dots;{\bf r}_{1'},{\bf r}_{2'},\dots $,
to denote the configuration of atoms in the gas, and 
particles $1,2,\dots$ have spin up and $1',2',\dots$ have spin down. 
The subscript FN denotes that the propagator is constrained such that the propagated 
wave function has the nodal surface of $\Psi_T({\bf R})$ at all $\tau$. 
The energy $E_T$ is adjusted to keep the norm of the wave function
constant. At large $\tau$, the evolved $\Psi(\tau, {\bf R})$ converges 
to the lowest energy state of the system having the
nodes of $\Psi_T({\bf R})$, and $E_T = \langle \Psi(\tau)| {\cal H}| \Psi(\tau) \rangle $.  
Without the fixed node constraint it will converge to the exact ground state, 
but unconstrained fermion Monte Carlo calculations become impractical due to uncontrolled 
growth in sampling errors.  This is the well known fermion sign problem. 
From now on, we assume that $\Psi_T({\bf R})$ is real and that $\tau$ is large enough to 
approximate the limit $\tau \rightarrow \infty$. 

Following Kalos \cite{kalos74}, the mixed expectation value
\beq
\langle {\cal{H}}(\tau) \rangle_{mixed}
=  \frac{\langle \Psi_T |{\cal H} |\Psi(\tau) \rangle}{\langle \Psi_T | \Psi(\tau) \rangle}
=  \frac{ \int d{\bf R} \Psi_T({\bf R}){\cal H} \Psi(\tau, {\bf R})}{\int d{\bf R} \Psi_T({\bf R}) \Psi(\tau, {\bf R})}
\eeq
is calculated using Monte Carlo sampling techniques. Since ${\cal H}$ commutes with the evolution operator we have
\beq
\langle {\cal H}(\tau) \rangle_{mixed}
=  \frac{\langle \Psi(\tau/2)|{\cal H} |\Psi(\tau/2) \rangle}{\langle \Psi(\tau/2) | \Psi(\tau/2) \rangle}
\equiv  \langle {\cal H} (\tau/2) \rangle~.
\eeq
The $ \langle {\cal{H}}(\tau) \rangle_{mixed}$ converges to the energy of the lowest 
enegy state with the nodal surface of $\Psi_T({\bf R})$. 
By the variational principle it is $ \geq $ the ground state energy $E_0$. 
The nodal surface of $\Psi_T( {\bf R})$ is varied to minimize $ \langle {\cal H}(\tau) \rangle_{mixed}$.
The minimum gives an accurate estimate of $E_0$ provided the variation is general enough.

\begin{figure}
\includegraphics[width=\columnwidth,clip]{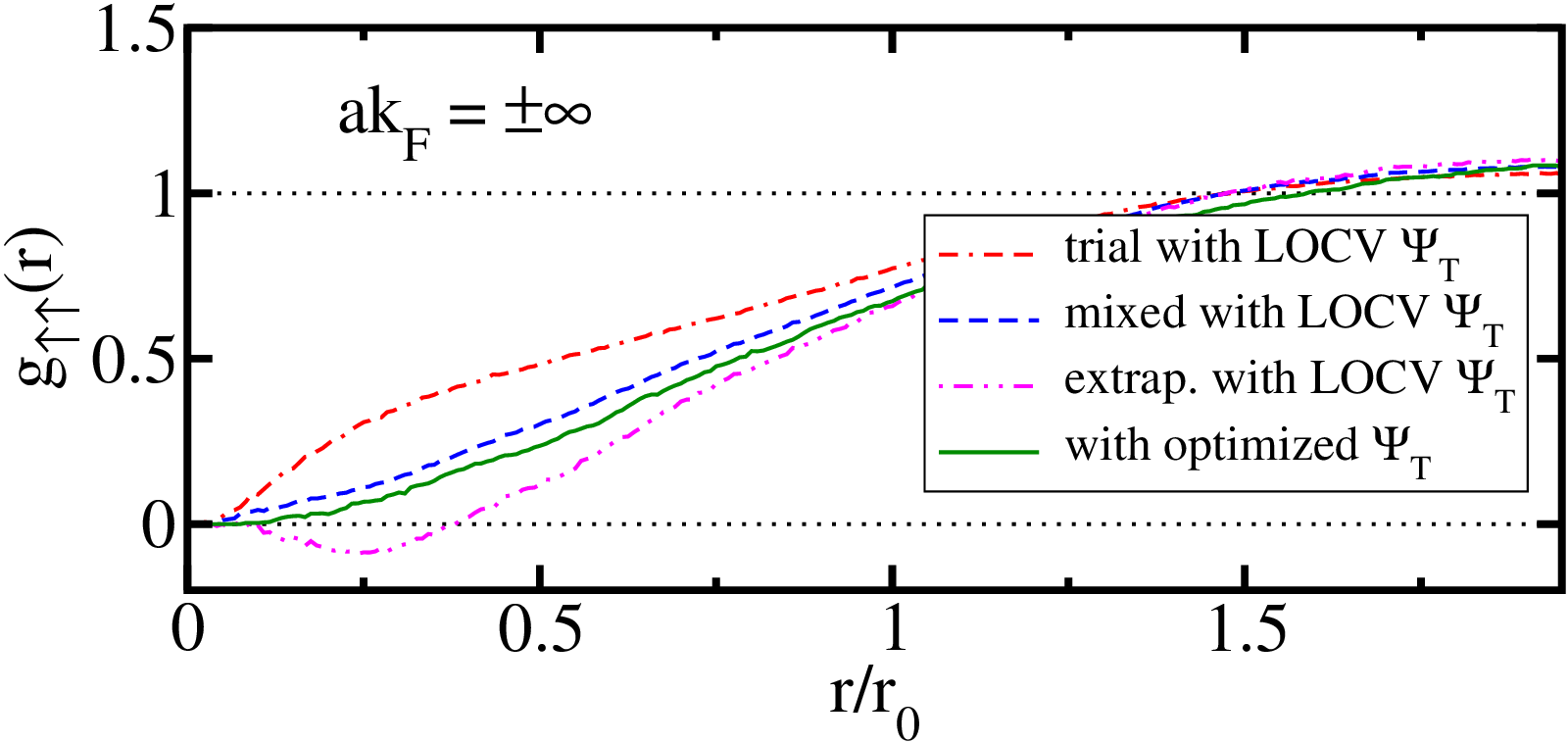}
\caption{ The trial, mixed and extrapolated $g_{\uparrow \uparrow}(r)$ with the LOCV 
$\Psi_T({\bf R})$ are compared with the trial $g_{\uparrow \uparrow}(r)$ with the $\Psi_{optim}({\bf R})$. 
Note that the mixed and trial pair distributions are the same for $\Psi_{optim}({\bf R})$. 
$r_0$ is given by $4\pi r_0^3\rho = 3$.}
\label{fig_guuinf}
\end{figure}

This procedure assures that the nodes of $\Psi_T({\bf R})$ are near optimum, i.e. close to the nodes of the exact
$\Psi_0({\bf R})$. However, the $\Psi_T({\bf R})$ itself can otherwise be very different from the $\Psi_0({\bf R})$.
For example, in references \cite{carlson2003} and \cite{chang2004} we use
\beq
\Psi_T({\bf R}) = \left[ \prod\limits_{i,j'}f_{\uparrow \downarrow}(r_{ij'}) \right] \Phi_{BCS}({\bf R}) ~,
\label{bcs_wf}  
\eeq
where $\Phi_{BCS}({\bf R})$ is a generalized Bardeen-Cooper-Schrieffer wave function, and its nodes are optimized by
minimizing $\langle {\cal H}(\tau) \rangle_{mixed}$. The $f_{\uparrow \downarrow}(r_{ij'})$ 
is a nodeless pair correlation function between spin up and down particles.
The $\langle {\cal H}(\tau) \rangle_{mixed}$ does not depend upon the
choice of $f_{\uparrow \downarrow}(r_{ij'})$ in the limit $\tau \rightarrow \infty$.  The 
$f_{\uparrow \downarrow}(r_{ij'})$ is used to reach this limit quickly, and to 
reduce the variance of the stochastic evaluation of $\langle {\cal H}(\tau) \rangle_{mixed}$.
Note that the commonly used Jastrow pair correlation function is also nodeless \cite{feenberg69}, but it acts 
between all pairs: $\uparrow \downarrow$, $\uparrow \uparrow$ and $\downarrow \downarrow$,  
and it is not useful in superfluid gases. 

Mixed expectation values of other observables, $\langle {\cal O}\rangle_{mixed}$,
can be easily calculated with GFMC, but they are more difficult to interprete when $[{\cal O}, {\cal H}] \ne 0$. 
If one assumes that $|\Psi(\tau) \rangle = |\Psi_T \rangle + |\delta \Psi \rangle $,
then the desired expectation value
\bea
\langle {\cal{O}}(\tau) \rangle 
& = &  \frac{\langle \Psi(\tau) | {\cal O} | \Psi(\tau) \rangle}{\langle \Psi(\tau)| \Psi(\tau)\rangle} \nonumber \\
& = &   2 \langle {\cal{O}}(\tau) \rangle_{mixed}  - \langle {\cal{O}} \rangle_{trial} \nonumber \\
& &  + \mbox{~terms of order }\delta\Psi^2~.
\eea
Here, the trial estimate $ \langle {\cal{O}} \rangle_{trial}  \equiv \langle \Psi_T|{\cal{O}}|\Psi_T \rangle /
\langle \Psi_T| \Psi_T \rangle$. When  $\delta \Psi$ is small, the extrapolation
\beq
\label{extrap}
\langle {\cal{O}}(\tau) \rangle_{extrap.} \approx 2 \langle {\cal{O}}(\tau) \rangle_{mixed}  
- \langle {\cal{O}} \rangle_{trial}~,
\eeq
can be used to estimate $\langle {\cal O } \rangle$.

\begin{figure}
\includegraphics[width=\columnwidth,clip]{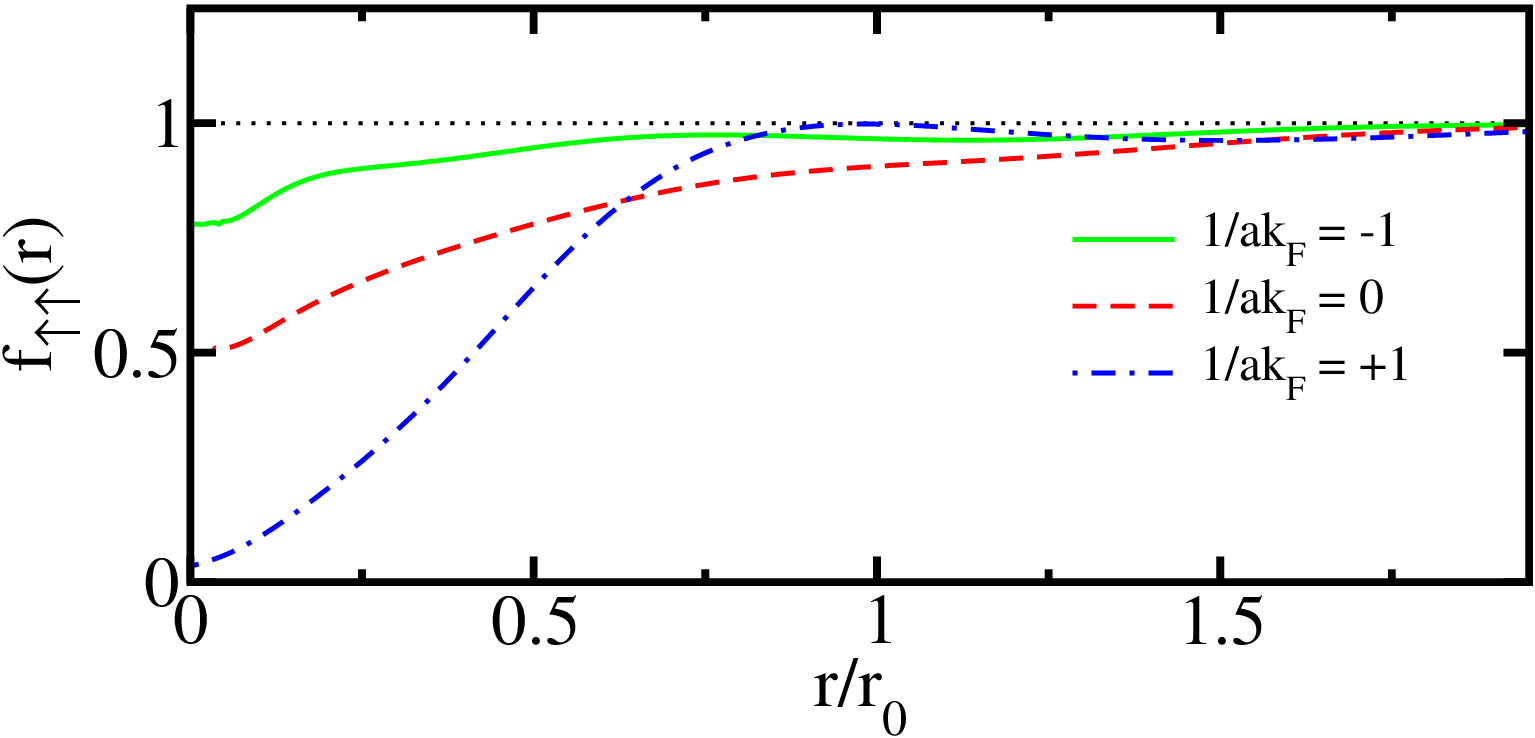}
\caption{ Optimized $f_{\uparrow \uparrow}(r)$ for different values of $ak_F$. }
\label{fig_fuu}
\end{figure}

\begin{figure}
\includegraphics[width=\columnwidth]{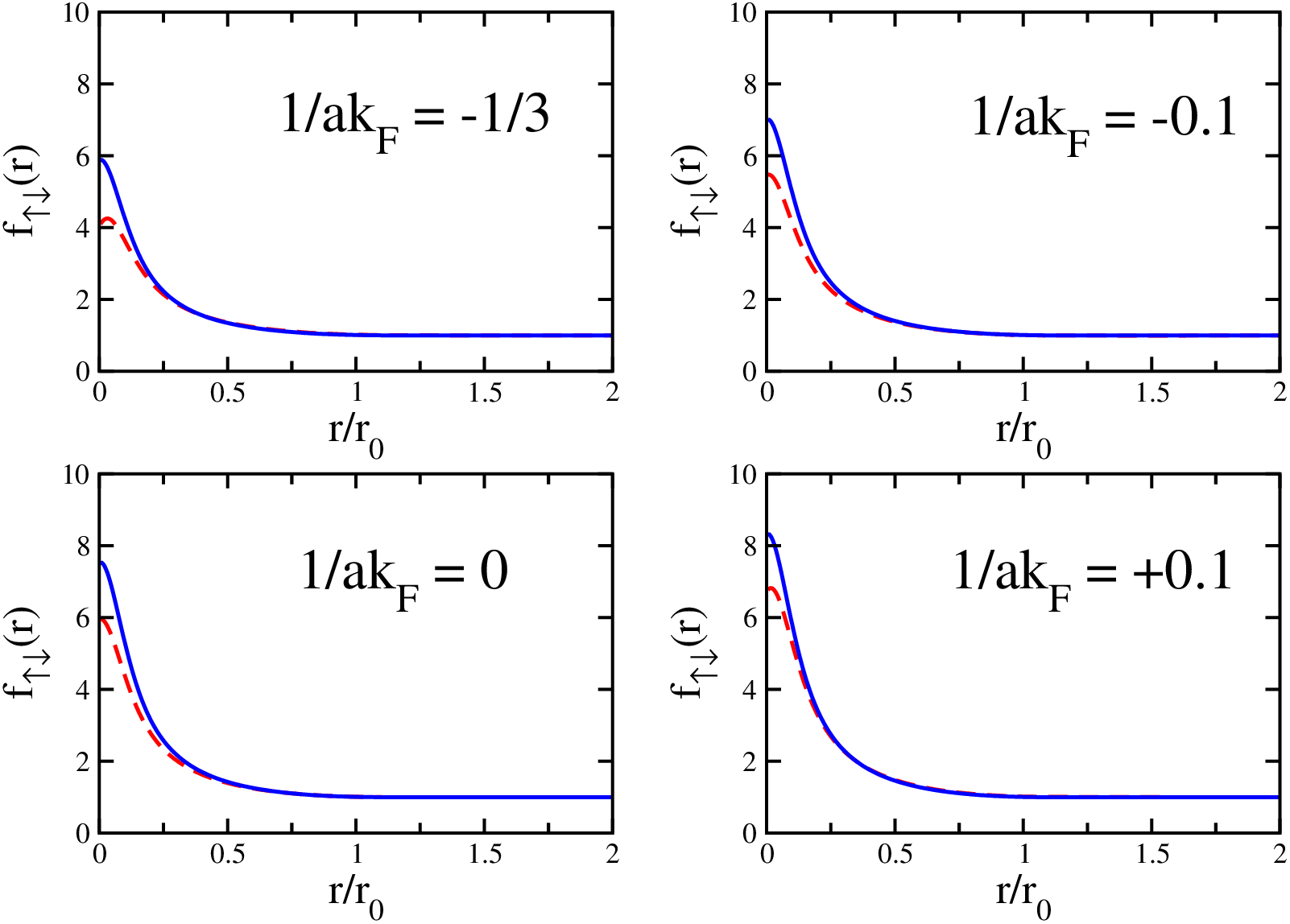}
\caption{ Optimized $f_{\uparrow \downarrow}(r)$ (dashed line) and LOCV $f_{\uparrow \downarrow}(r)$ (continuous line)
 for different values of $ak_F$.}
\label{fig_fud}
\end{figure}

However, in the strongly interacting regime the $\delta\Psi$ is not necessarily  
small, and this extrapolation may not be
valid. For example, we consider the pair distribution function
$g_{\uparrow \uparrow}(r)$ between parallel spin particles in the 
$ak_F \rightarrow \pm\infty$ limit. The mixed, trial and
extrapolated values of $g_{\uparrow \uparrow}(r)$ obtained from the $\Psi_T({\bf R})$ of 
Ref. \cite{carlson2003} and \cite{chang2004}
are shown in Fig. \ref{fig_guuinf}. At small $r$, the extrapolated 
$g_{\uparrow \uparrow}(r) < 0 $ indicating invalidity.
These and all the other results presented in this work are obtained from Monte Carlo computations 
using 14 particles in a cubic periodic box.  As discussed in Ref.~\cite{carlson2003} and \cite{chang2004},
 a periodic box with 14 particles provides a fair approximation to the uniform gas. 
The cosh $v(r_{ij'})$ with $\mu r_0=12$ is used \cite{chang2004} to approximate the interaction between 
spin $\uparrow \downarrow $ pairs.

\begin{table}
\begin{center}
\caption{Summary of the results in units of $E_{FG}$}
\begin{tabular}{|c|c|c|c|c|c|} \hline
$ ak_F $ & $\langle{\cal{H}}\rangle_{mixed}$ & $\langle{\cal{H}}\rangle_{trial}$ & $\langle{\cal{H}}\rangle_{optim}$ 
& $\langle{\cal{V}}\rangle_{mixed}$ & $\langle{\cal{V}}\rangle_{optim}$\\
\hline
\hline
-1  & 0.792(4) & 0.818(3) & 0.808(4) & -0.55(3) & -0.54(3) \\
-3  & 0.635(6) & 0.85(3) & 0.70(2)   & -1.8(1) & -2.0(1)\\
-10  & 0.494(7) & 0.68(3) & 0.53(1)  & -3.5(2) & -3.2(1)\\
$\infty$ &  0.414(5) & 0.62(3) & 0.46(1) & -3.9(1) & -4.0(2)\\
10  &  0.32(1) & 0.57(6) & 0.39(1)  &  -4.8(1) & -5.0(1)\\
3  &  -0.00(1) & 0.4(1) & 0.11(3)   &  -7.0(1) & -7.3(3)\\
2  &  -0.34(2) & 0.2(1) & -0.18(3)  &  -9.2(4) & -9.2(3)\\
1  &  -2.37(3) & -0.1(1) & -2.01(3) &   -19.0(4) & -18.0(6) \\
\hline 
\end{tabular}
\label{table_res}
\end{center}
\end{table}

In principle, the pair correlation functions, $f_{\uparrow \downarrow}(r)$ and $f_{\uparrow \uparrow}(r)
= f_{\downarrow \downarrow}(r)$ in $\Psi_T({\bf R})$ can be obtained by minimizing the trial energy
$\langle {\cal H}\rangle_{trial}$.  However, this variational problem has been approximately
treated in most quantum Monte Carlo calculations.  In Ref. \cite{carlson2003,chang2004} a 
simple and crude method called LOCV, based on constrained minimization of the leading two-body cluster contribution to
$\langle {\cal H}\rangle_{trial}$ \cite{kevin77} is used. In this method,
$f_{\uparrow \uparrow}(r) = f_{\downarrow \downarrow}(r) = 1$, and $f_{\uparrow \downarrow}(r)$ satisfies 
the two-body Schr\"odinger equation
\beq
-\frac{\hbar^2}{m}\nabla^2f_{\uparrow \downarrow}(r) + v(r) f_{\uparrow \downarrow}(r) = \lambda f_{\uparrow \downarrow}(r)~,
\label{locv_eqn}
\eeq
at $r<d$. The boundary conditions are: $f_{\uparrow \downarrow}(r \geq d) = 1$ and $f'_{\uparrow \downarrow}(r=d)=0$. The
healing distance $d$ serves as the variational parameter.
The trial energies obtained with variational Monte Carlo (VMC) calculations using the optimum healing distance 
$d$ are compared with the FN-GFMC $\langle {\cal H} \rangle_{mixed}$ in Table \ref{table_res}.  Both  
calculations use the optimum $\Phi_{BCS}({\bf R})$ found by minimizing the $\langle {\cal H} \rangle_{mixed}$ 
in Ref. \cite{chang2004}.  The trial energies are well above $\langle {\cal H} \rangle_{mixed}$, 
particularly at $1/ak_F \geq 0$. This shows that the LOCV pair correlation functions are 
far from those in the exact $\Psi_0$ in the strongly interacting regime.

  Here we present a new and simple method to optimize 
the pair correlation functions in the trial wave function using GFMC results.  
The optimized trial wave functions, denoted by $\Psi_{optim}({\bf R})$  are presumably close enough to $\Psi_0({\bf R})$ 
so that $\delta \Psi $ is small and Eq.~\ref{extrap} provides a fair approximation. The method can be improved
 and $\delta\Psi$ can be further decreased by including higher order correlations corresponding to triplet, quadruplet, etc. However, in the present work
we consider only pair correlations for $\Psi_{optim}({\bf R})$ . 

The trial pair distribution functions can be expressed as \cite{feenberg69}
\beq
\langle g_x(r) \rangle_{trial} = f_x^2(r)~ t_x(r, f_{\uparrow \uparrow}, f_{\uparrow \downarrow},\Phi_{BCS}({\bf R}),\rho)~,
\label{gnf}
\eeq
where $x$ can be $\uparrow \uparrow$ or $\uparrow \downarrow$ and $t_x$ is a complicated function of $r$,
$f_{\uparrow \uparrow}, f_{\uparrow \downarrow}$, $\Phi_{BCS}({\bf R})$ and gas density $\rho$. It is difficult to
calculate it exactly except by numerical
methods. However, $t_x(r)$ contains many-body integrals, and is a relatively smooth function of $r$.

Our method to optimize $f_x(r)$ using GFMC is iterative. 
Let $\langle g_x^{(n)}(r)\rangle_{mixed}$ and $\langle g_x^{(n)}(r)\rangle_{trial}$ be obtained from the n-th 
trial $f_x^{(n)}(r)$ using the optimum $\Phi_{BCS}$ which does not depend on $f_x(r)$. 
We start with the LOCV approximation providing the $f_x^{(1)}(r)$, but one could start with 
any other choice of $f_x^{(1)}$ and converge to the same $\Psi_{optim}({\bf R})$ . 
The next improved $f_x^{(2)}(r)$ is chosen as
\beq
f_{x}^{(2)}(r)  = f_{x}^{(1)}(r) \sqrt{\frac{ \langle g_x^{(1)}(r)\rangle_{mixed} }
{ \langle g_x^{(1)}(r)\rangle_{trial}}}~.
\label{opt_step}
\eeq
If the difference between $f_x^{(1)}(r)$ and $f_x^{(2)}(r)$ is small, 
we can assume that the $t_x(r)$ functions do not change
much. In this case $ \langle g_x^{(2)}(r)\rangle_{trial} \approx \langle g_x^{(1)}(r)\rangle_{mixed}$.
Otherwise, by iterating this process one easily converges to an $f_x^{(n)}(r)$ such that
\beq
\langle g_x^{(n)}(r)\rangle_{trial} \approx \langle g_x^{(n)}(r)\rangle_{mixed}~.
\label{converged}
\eeq
Usually, the convergence within statistical errors can be reached within 3 $\sim$ 4 iterations and it
doesn't seem to depend on the strength of the interaction.
In practice, $\langle g_x(r) \rangle_{mixed}$ and $\langle g_x(r)\rangle_{trial}$
have Monte Carlo sampling errors. We approximate the square root of their ratio 
(Eq.~\ref{opt_step}) by a smooth function of $r$ chosen as 
$\cos(p_1 r+p_2)e^{-r/p_3}+1$, and vary the parameters $p_{1-3}$ to best fit the Monte Carlo values.
One iteration step typically takes about 10 hours in a Pentium 3.0 GHz based workstation.

The VMC energies with the $\Psi_{optim}({\bf R})$  are much closer to the FN-GFMC energies (Table \ref{table_res}).
In principle, the optimization of $f_x(r)$ should have no effect on the FN-GFMC $\langle {\cal H} \rangle_{mixed}$; in practice the 
$\langle {\cal H} \rangle_{mixed}$ seems to get lowered by $\sim 2 \pm 1$ \% after optimization presumably because the limit $\tau \rightarrow \infty$ is easier to reach 
with the $\Psi_{optim}({\bf R})$. 
The effects of the optimization are also seen in the reduced error bars of the
energy estimates: $\delta\langle {\cal H} \rangle_{optim} \lesssim \delta\langle {\cal H} \rangle_{trial}$ (Table \ref{table_res})
for the same number of Monte Carlo samples.
In addition, the $E_T$ (Eq. \ref{t_evol}), which typically has larger fluctuations,
becomes indistinguishable from $ \langle {\cal H} \rangle_{mixed}$. 

The pair distribution functions $\langle g_x(r)\rangle_{mixed}$ are
determined by the many body probability distribution given by $\Psi_{optim}({\bf R})\Psi(\tau,{\bf R})$,
while the $\langle g_x(r)\rangle_{optim}$ are for $|\Psi_{optim}({\bf R})|^2$. 
Note that $\Psi_{optim}({\bf R})\Psi(\tau,{\bf R}) \approx \Psi_{optim}({\bf R})\Psi_0({\bf R}) \geq 0$ since
the nodes of $\Psi_{optim}({\bf R})$ have been varied to match those of $\Psi_0({\bf R})$.

 Extending the above method, if we can match the mixed and optimized trial distributions for all, pair,
 triplet, quadruplet, $\dots$ distribution functions, then we can assume that
$\Psi_{optim}({\bf R}) = \Psi_0({\bf R})$. However, here we approximate the exact $\Psi_0({\bf R})$ by $\Psi_{optim}({\bf R})$ using $\Phi_{BCS}({\bf R})$ and pair
correlation functions only
\bea
\Psi_{optim}({\bf R}) &=& \prod\limits_{i,j'}f_{\uparrow \downarrow}^{optim}(r_{ij'})
 \prod\limits_{i<j}f_{\uparrow \uparrow}^{optim}(r_{ij}) \nonumber \\
  & &\times \prod\limits_{i'<j'}f_{\downarrow \downarrow}^{optim}(r_{i'j'})  \Phi_{BCS}({\bf R})~.
 \label{new_bcs_wf}
\eea
The validity of Eq. \ref{converged} ensures that the present optimization method will 
converge to $f_x^{(n)} \rightarrow f_x^{optim}(r)$ and thus $\Psi_{optim}({\bf R})$ is as
close to $\Psi_0({\bf R})$ as its form (Eq. \ref{new_bcs_wf}) allows.
 If higher order correlations have negligible effects on the wave functuion, we should expect 
\beq
\langle {\cal H} \rangle_{optim} \approx \langle {\cal H} \rangle_{mixed} \approx E_0~.
\eeq
However, in the interesting regime of $ak_F \sim \infty$, Table \ref{table_res} shows that the 
$\langle {\cal H} \rangle_{optim}$ is larger than the $\langle {\cal H} \rangle_{mixed}$ by 
$\sim$ 10 \%.  This suggests that the form of the present $\Psi_{optim}({\bf R})$ is not sufficiently general.
An improved approximation could be obtained by including products of triplet 
correlations $F_{P}(r_{ij}, r_{jk},r_{ki})$ for $\uparrow \uparrow \uparrow$ and  
$\downarrow \downarrow \downarrow$, and $F_{M}(r_{ij}, r_{jk'}, r_{k'i})$ for $\uparrow \uparrow \downarrow$ and 
$\downarrow \downarrow \uparrow$ triplets in the wave function.  We believe that the present method can be 
generalize to determine the optimal forms of three-body correlations by making
\beq
g_{3,x}(r_{ij},r_{jk},r_{ki})_{mixed} = g_{3,x}(r_{ij},r_{jk},r_{ki})_{optim}~,
\eeq
where $g_3$ denotes three particle distribution functions.  The true $\Psi_0$ can also have 
backflow correlations \cite{kevin81}; however, they change the nodal surface
and have to be optimized by minimizing $\langle {\cal H} \rangle_{mixed}$.

The main difference between the optimum pair correlations and those of Ref. \cite{carlson2003}
and \cite{chang2004} is in $f_{\uparrow \uparrow}(r)$.
In LOCV we have $f_{\uparrow \uparrow}(r) = 1$, because in two-body clusters 
there is no interaction between
parallel spin particles in dilute Fermi gases. However, many body effects generate an effective repulsion
between parallel spin particles and the optimum $f_{\uparrow \uparrow}(r)$ is $<1$
at $r \lesssim 1.5 r_0$ as shown in Fig. \ref{fig_fuu}.

The optimum and LOCV $f_{\uparrow \downarrow}(r)$ are generated by the
strong two body attraction in $\uparrow \downarrow$ pairs and have qualitatively similar shapes
(Fig. \ref{fig_fud}). For $1/ak_F << -1$, the LOCV $f_{\uparrow \downarrow}(r)$ is near optimum.
For stronger interactions, it is larger than the optimum at $r \sim 0$ (Fig. \ref{fig_fud}). 

\begin{figure}
\includegraphics[width=\columnwidth]{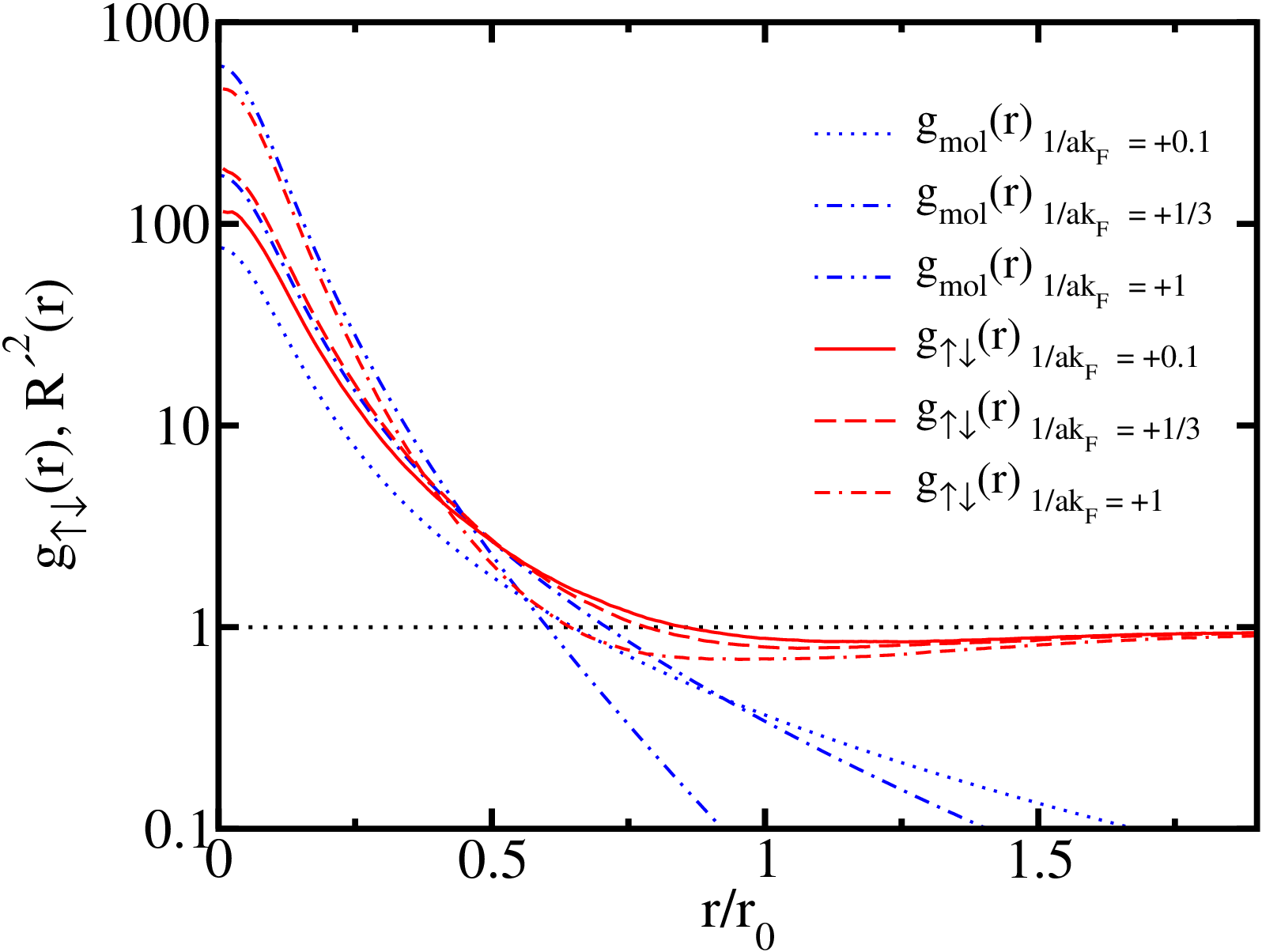}
\caption{ Comparison of the pair distribution function $g_{\uparrow \downarrow}(r)$ with the radial
probability distribution of the molecule in the $a>0$ regime. }
\label{fig_grsq}
\end{figure}

The expectation value of the potential energy, $\langle {\cal V} \rangle = \langle \sum_{i,j'} v(r_{ij'}) \rangle$ can easily be
calculated from the GFMC (mixed) and VMC distributions using $\Psi_{optim}$. 
The calculated values of the potential energy are given in the last two columns of Table \ref{table_res}. 
Apart from statistical fluctuations, the mixed and the optimum pair distribution functions are the same. 
Therefore no extrapolation, such as in Eq. \ref{extrap}, is necessary for calculating $\langle {\cal V} \rangle$
using $\Psi_{optim}({\bf R})$.

Only when $1/ak_F > 0$, we can have bound states with normalized radial wave functions $R(r)$. We
define $R'(r) = \sqrt{\frac{2}{\rho}}R(r)$ so that $g_{mol}(r_{\uparrow \downarrow}) \equiv R'^2(r)$ is normalized analogous to 
$g_{\uparrow \downarrow}(r)$, and the two are compared in Fig. \ref{fig_grsq} 
($g_{\uparrow \downarrow}(r)$ is normalized such that $g_{\uparrow \downarrow}(r) \rightarrow 1$ for 
$ r \rightarrow \infty$).
When $1/ak_F \rightarrow 0^+$, we know that $g_{mol}(r) = R'^2(r) \rightarrow 0$ (infinite pair size),
but $g_{\uparrow \downarrow}(r) \gtrsim 1$. So $g_{mol}(r)$ and $g_{\uparrow \downarrow}(r)$ are qualitatively different
when $a$ is large. However, when the interaction is stronger and $a$ becomes positive and small we expect a gas of molecules
in which $g_{\uparrow \downarrow}(r) \sim g_{mol}(r)$ at $r < $ the size of the molecule. Fig.~\ref{fig_grsq} shows that
the superfluid may well be approximated by a gas of molecules with BEC for $1/ak_F \gtrsim 1/3$.
 The molecule size for $1/ak_F = 1/3$, is $1.21 r_0$ and for $1/ak_F = 1$, it is $0.38 r_0$. Due to
 the many body effects in $g_{\uparrow \downarrow}(r)$, it is meaningless to compare beyond these
 distances. In fact, for all $ak_F>0$, the $g_{mol}(r) \rightarrow 0$ while $g_{\uparrow \downarrow}(r) \rightarrow 1$ 
 as $r \rightarrow \infty$.
 
In conclusion, the proposed method allows us to optimize separately the BCS and the
pair correlations in dilute Fermi gases. The BCS and
$f_{\uparrow \downarrow}(r)$ correlations are most important, however in the strong interaction regime,
the $f_{\uparrow \uparrow}(r)$ can not be neglected. 

The studies of momentum distributions and density matrices of the superfluid gas may now be
possible using the optimum $\Psi_T({\bf R})$, and are in progress.

This work is partly supported by the U.S. National Science Foundation via grant PHY-03-55014.

\end{document}